\documentclass[aps,prc,superscriptaddress,twocolumn]{revtex4-2}

\usepackage{graphicx}
\usepackage{dcolumn}
\usepackage{bm}
\usepackage{amsmath}
\usepackage{amssymb}
\usepackage{isotope}
\usepackage{color}
\usepackage{flushend}
\usepackage[export]{adjustbox}

\begin{document}

\title{Search for the origin of wobbling motion in the $ \boldsymbol{A \approx 130} $ region: The case of \textsuperscript{131}Xe}

\author{S.~Chakraborty}
\email{saikat.c@vecc.gov.in}
\affiliation{Variable Energy Cyclotron Centre, Kolkata - 700064, India}
\author{S.~Bhattacharyya}
\email{sarmi@vecc.gov.in}
\affiliation{Variable Energy Cyclotron Centre, Kolkata - 700064, India}
\affiliation{Homi Bhabha National Institute, Mumbai - 400094, India}
\author{R.~Banik}
\affiliation{Institute of Engineering and Management, Kolkata - 700091, India}
\author{Soumik~Bhattacharya}
\altaffiliation{Presently at: Florida State University, USA}
\affiliation{Variable Energy Cyclotron Centre, Kolkata - 700064, India}
\author{G.~Mukherjee}
\author{C.~Bhattacharya}
\affiliation{Variable Energy Cyclotron Centre, Kolkata - 700064, India}
\affiliation{Homi Bhabha National Institute, Mumbai - 400094, India}
\author{S.~Biswas}
\altaffiliation{Presently at: Paul Scherrer Institute, Switzerland}
\affiliation{Grand Accélérateur National d’Ions Lourds (GANIL), CEA/DRF-CNRS/IN2P3, BP 55027, F-14076 Caen Cedex - 5, France}
\author{S.~Rajbanshi}
\affiliation{Department of Physics, Presidency University, Kolkata - 700073, India}
\author{Shabir~Dar}
\affiliation{Variable Energy Cyclotron Centre, Kolkata - 700064, India}
\affiliation{Homi Bhabha National Institute, Mumbai - 400094, India}
\author{S.~Nandi}
\altaffiliation{Presently at: Argonne National Laboratory, USA}
\affiliation{Variable Energy Cyclotron Centre, Kolkata - 700064, India}
\affiliation{Homi Bhabha National Institute, Mumbai - 400094, India}
\author{Sajad~Ali}
\affiliation{Department of Physics, Government General Degree College at Pedong, Kalimpong - 734311, India}
\author{S.~Chatterjee}
\author{S.~Das}
\affiliation{UGC-DAE Consortium for Scientific Research, Kolkata Centre, Kolkata - 700098, India}
\author{S.~Das~Gupta}
\affiliation{Department of Physics, Victoria Institution (College), Kolkata - 700009, India}
\author{S.~S.~Ghugre}
\affiliation{UGC-DAE Consortium for Scientific Research, Kolkata Centre, Kolkata - 700098, India}
\author{A.~Goswami}
\thanks{Deceased}
\affiliation{Saha Institute of Nuclear Physics, Kolkata - 700064, India}
\affiliation{Homi Bhabha National Institute, Mumbai - 400094, India}
\thanks{deceased}
\author{A.~Lemasson}
\affiliation{Grand Accélérateur National d’Ions Lourds (GANIL), CEA/DRF-CNRS/IN2P3, BP 55027, F-14076 Caen Cedex - 5, France}
\author{Debasish~Mondal}
\affiliation{Variable Energy Cyclotron Centre, Kolkata - 700064, India}
\author{S.~Mukhopadhyay}
\affiliation{Variable Energy Cyclotron Centre, Kolkata - 700064, India}
\affiliation{Homi Bhabha National Institute, Mumbai - 400094, India}
\author{A.~Navin}
\affiliation{Grand Accélérateur National d’Ions Lourds (GANIL), CEA/DRF-CNRS/IN2P3, BP 55027, F-14076 Caen Cedex - 5, France}
\author{H.~Pai}
\affiliation{Extreme Light Infrastructure - Nuclear Physics (ELI-NP), IFIN-HH, Bucharest-Magurele - 077126, Romania}
\author{Surajit~Pal}
\affiliation{Variable Energy Cyclotron Centre, Kolkata - 700064, India}
\author{Deepak~Pandit}
\affiliation{Variable Energy Cyclotron Centre, Kolkata - 700064, India}
\affiliation{Homi Bhabha National Institute, Mumbai - 400094, India}
\author{R.~Raut}
\affiliation{UGC-DAE Consortium for Scientific Research, Kolkata Centre, Kolkata - 700098, India}
\author{Prithwijita~Ray}
\affiliation{Department of Physics, Acharya Brojendra Nath Seal College, Coochbehar - 736101, India}
\author{M.~Rejmund}
\affiliation{Grand Accélérateur National d’Ions Lourds (GANIL), CEA/DRF-CNRS/IN2P3, BP 55027, F-14076 Caen Cedex - 5, France}
\author{S.~Samanta}
\altaffiliation{Presently at: University of Genoa, Italy}
\affiliation{UGC-DAE Consortium for Scientific Research, Kolkata Centre, Kolkata - 700098, India}

\begin{abstract}
In-beam $ \gamma $-ray spectroscopy of \isotope[131]Xe has been carried out to study the structure of the intruder $ \nu h_{11/2} $ band. Excited states were populated via an $ \alpha $-induced fusion-evaporation reaction at E$ _{\alpha} = 38 $ MeV. Inspection of $ \gamma \gamma $-coincidence data resulted in the identification of a new rotational sequence. Based on the systematics of excitation energy, assigned spin-parity, decay pattern, and the electromagnetic character of the inter-band $ \Delta I = 1 $ $ \gamma $-transitions, this sequence is proposed as the unfavoured signature partner of the $ \nu h_{11/2} $ band. The structure of this band is further illuminated in the light of the triaxial particle rotor model (TPRM). The possibility of wobbling excitation in $ N = 77 $ Xe-Ba-Ce isotones has been explored in a systematic manner.
\end{abstract}

\maketitle

\section{Introduction}

Rotational motion is a typical collective mode of excitation in atomic nuclei \cite{Rotational_motion}. It originates to restore the rotational symmetry broken by nuclear deformation. The wave function of an axially symmetric (prolate or oblate) nucleus is invariant with respect to a rotation by an angle of $ 180^{\circ} $ about an axis perpendicular to its symmetry axis ($ \mathcal{R} $). The quantum number associated with the $ \mathcal{R} $ operator is known as signature ($ \alpha $) \cite{bohr&mottelson-II}. The even and odd spin sequences of a rotational band in even-\textit{A} nuclei correspond to $ \alpha = 0,~1 $, respectively. Likewise, the $ I = \frac{1}{2}, \frac{5}{2}, \frac{9}{2}, ... $ and $ I = \frac{3}{2}, \frac{7}{2}, \frac{11}{2}, ... $ sequences in an odd-\textit{A} nucleus correspond to $ \alpha = \pm 1/2 $. The signature-dependent splitting in energy is known as signature splitting $ S(I) $ and can readily be extracted from the experimentally deduced level energies. The magnitude of $ S(I) $ has a distinct \textit{K}-dependence (\textit{K} is the projection of total angular momentum on the symmetry axis) \cite{EM_Donau}. For instance, in an axially symmetric nucleus, a rotational band with a high-\textit{K} (low-\textit{K}) configuration is predicted to exhibit a small (large) signature splitting \cite{119I}. However, in the triaxially deformed nuclei, the quantity \textit{K} no longer remains conserved and hence the band structures in these nuclei have mixed configurations of wave functions with different \textit{K} values. As a consequence, a rotational phenomenon like signature splitting is found to appear in a different way than expected \cite{Xe_triaxiality_NPA}. Thus, the quantity $ S(I) $ was proposed to quantify the degree of triaxiality in atomic nuclei \cite{Re_PRM}. 

The rotational motion of a triaxially deformed nucleus can be realised by observing a pair of chiral doublet bands or a wobbling band or a $ \gamma $-band \cite{bohr&mottelson-II,chiral_TAC,wobbling_QTR,gamma-band}. A large number of experimental signatures in favour of triaxial nuclear shapes has been found in the $ A \approx 130 $ region, mainly, due to the presence of the unique parity shape driving $  h_{11/2} $ orbital. Among these, the occurrence of wobbling bands at low angular momentum in normal-deformed $ \gamma $-soft nuclei has drawn a lot of attention in the recent past. The rotational properties, like, moments of inertia, alignments \textit{etc.} and the decay pattern of these bands are quite similar to that of the unfavoured signature partner band. But, the electromagnetic properties of the decay between favoured and unfavoured signature partner bands were found to be different from the electromagnetic properties of the decay between two successive phonon wobbling bands. In the case of signature partner band, the $ \Delta I = 1 $ transitions are primarily $ M 1 $ in nature. On the other hand, such $ \Delta I = 1 $ $ \gamma $-rays between two consecutive phonon wobbling bands are expected to have large  $ E2 $ component. Thus, the structure of the $ \alpha = + 1/2 $ partner of the intruder $ h_{11/2} $ bands in odd-\textit{A} triaxial nuclei has become a topic of current interest. In recent studies, based on the large $ \delta $ value of the connecting $ \Delta I = 1 $ $ \gamma $-transitions, the so-called unfavoured signature partner of the $ \pi h_{11/2} $  bands in \isotope[135]Pr \cite{135Pr_PRL}, \isotope[133]La \cite{133La_EPJA} and the $ \nu h_{11/2} $ bands in \isotope[133]Ba \cite{133Ba_PLB}, \isotope[127]Xe \cite{127Xe_PLB}, \isotope[105]Pd \cite{105Pd_PRL} have been reinterpreted in terms of wobbling excitation. In this context, it is also worth mentioning that in these nuclei (except in \isotope[133]La) apart from the wobbling band the unfavoured signature partner band is also identified.

Since Xe isotopes are well known to have triaxial shapes \cite{Xe_triaxiality_NPA}, searching for the wobbling bands even in the heavier Xe nuclei would be promising. However, very recently most of these wobbling assignments are questioned from theoretical and/or experimental perspectives \cite{antiwobbling_IBM,135Pr_antiwobbling}. Hence, besides searching for the wobbling bands, proper identification of the unfavoured signature partner band is essential to understand the nuclear structure of triaxial nuclei and to avoid any misinterpretation of the wobbling phenomenon. 
\begin{figure}[!t]
\centering
\includegraphics[height=0.9\columnwidth, angle=270]{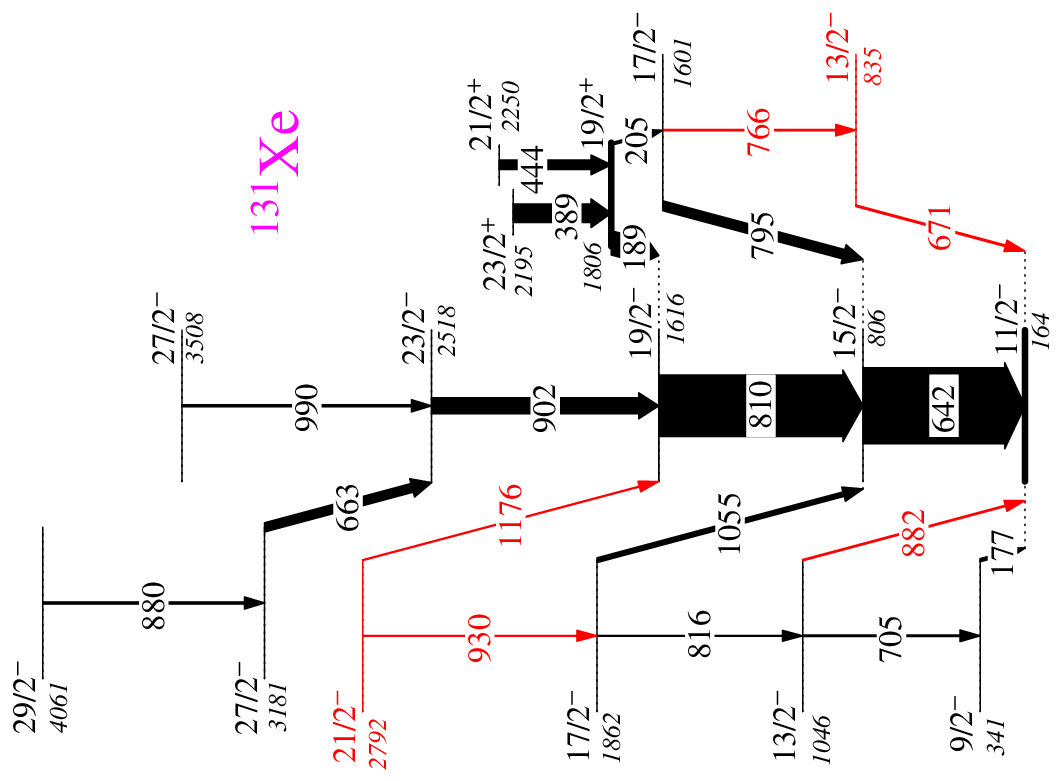}
\caption{Partial level scheme of \isotope[131]Xe, showing the negative parity levels and decaying $ \gamma $ rays of present interest. New experimental information are marked in red colour.}
\label{lev-sch}
\end{figure}
Among the more neutron-rich Xe isotopes, the experimental information on the low-lying yrast and non-yrast states in \isotope[131]Xe were extracted primarily from lighter-ion-induced fusion-evaporation reactions \cite{131Xe_NPA,131Xe-JPG,131Xe_JPG,131Xe_PS}. Later, the yrast negative parity band was extended up to $ I^{\pi} = (35/2^{-}) $ using two independent heavy-ion induced reactions \cite{131Xe-PRC}. However, only a limited spectroscopic information was reported on this nucleus from all of these studies. Recently, a detailed spectroscopic investigation of the high spin states in \isotope[131]Xe was carried out using an $ \alpha $-induced reaction and that significantly extended the level scheme of this nucleus \cite{131Xe_PRC}. But, unlike the lighter Xe isotopes, the unfavoured signature partner of the $ \nu h_{11/2} $ band was unobserved. As the odd-quasineutron in this nucleus is expected to occupy the higher-$ \Omega $ orbitals, the unfavoured partner of the $ \nu h_{11/2} $ band is expected to be observed. It is also evident from the recent observation of the unfavoured partner of the $ \nu h_{11/2} $ band in $ N = 77 $ \isotope[133]Ba isotone \cite{133Ba_PLB}. Therefore, the aim of the present work is to search for the expected but heretofore unreported unfavoured signature partner of the $ \nu h_{11/2} $ band and/or wobbling band in $ ^{131}_{54} $Xe$ _{77} $ nucleus from a re-analysis of the data reported in Ref.~\cite{131Xe_PRC}.

\begin{figure}[!t]
\centering
\includegraphics[width=\columnwidth]{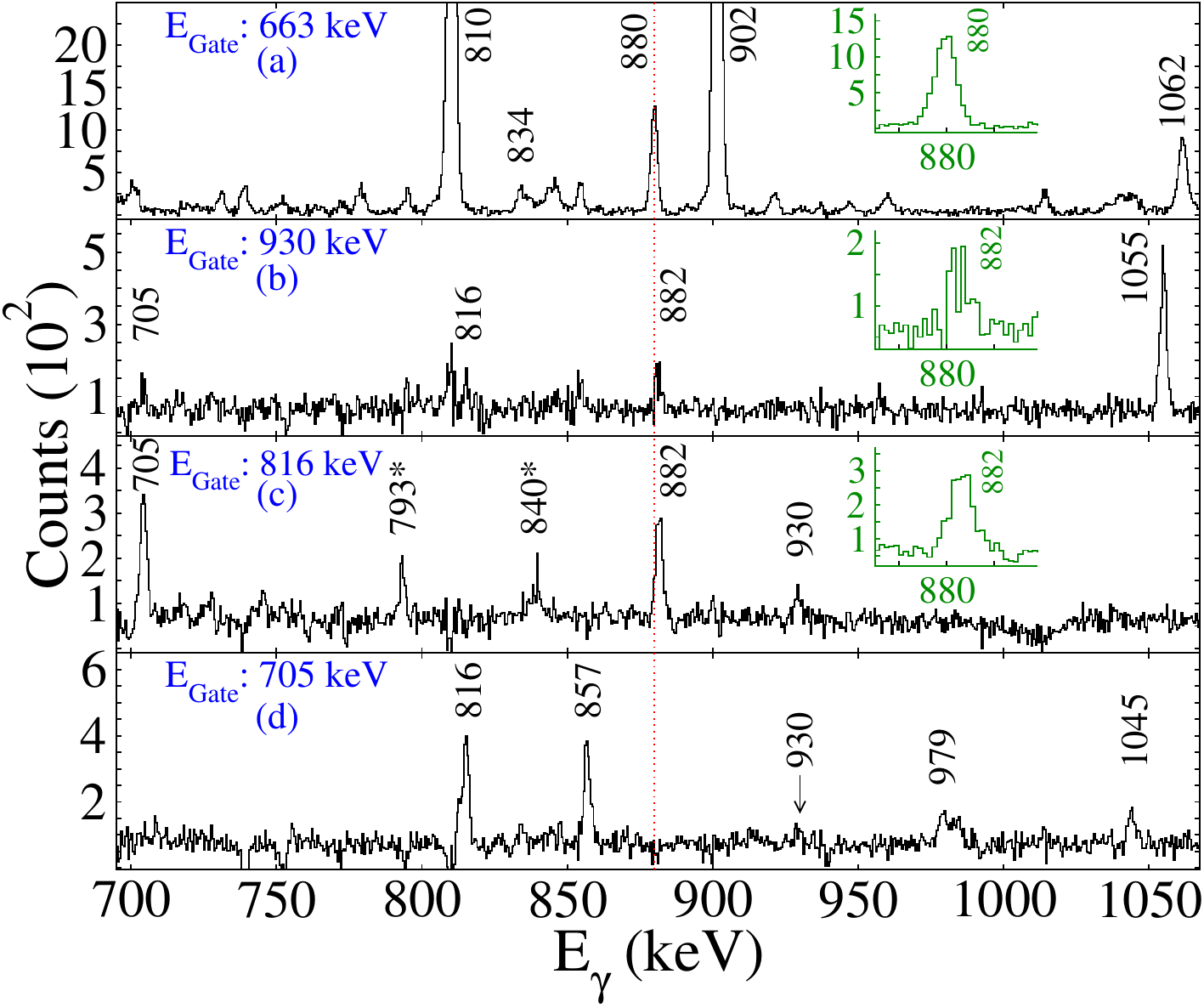}
\caption{Coincidence $ \gamma $ spectra, gated by (a) 663, (b) 930, (c) 816 and (d) 705 keV (top to bottom), to show the $ \gamma $ rays of present interest in \isotope[131]Xe. The $ \gamma $ rays which are marked here but not shown in the partial level scheme of \isotope[131]Xe are already reported in Ref.~\cite{131Xe_PRC}. The asterisk ($ \ast $) marked $ \gamma $ rays are belonging to \isotope[130]Te. The insets show the centroid difference of the 880 and 882 keV doublet transitions in the 663 and 816, 930 keV energy gates, respectively.}
\label{spec}
\end{figure}
\begin{figure}[!b]
\centering
\includegraphics[width=\columnwidth]{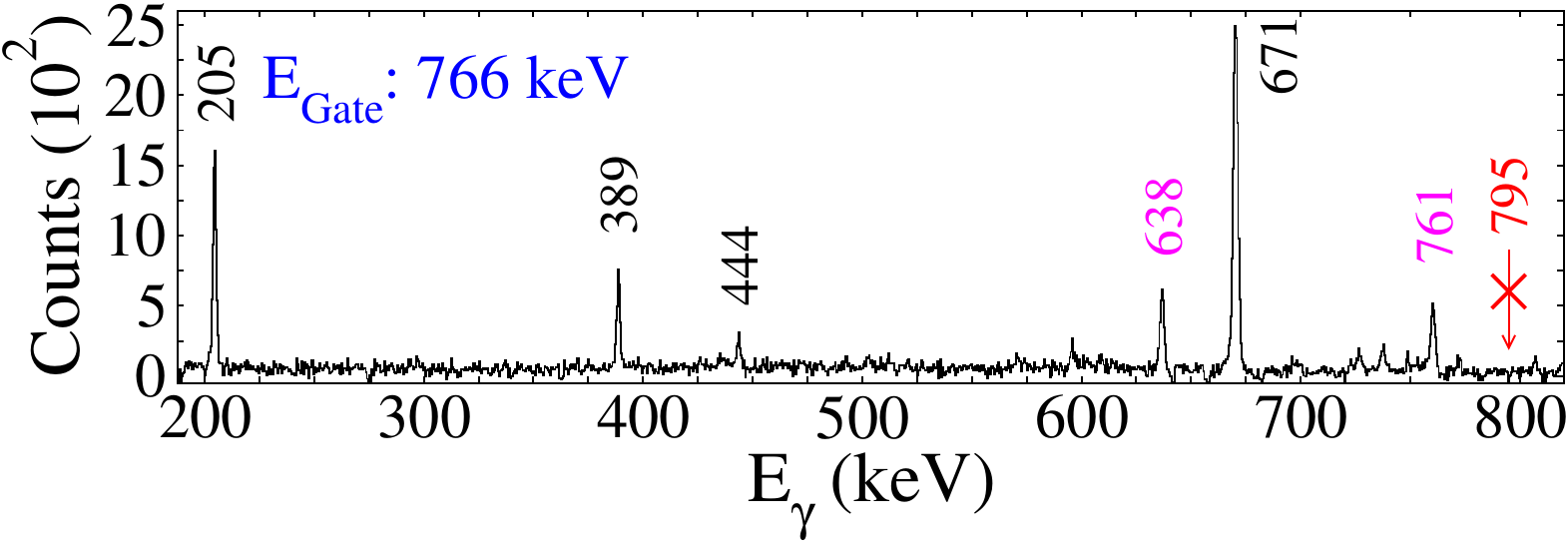}
\caption{Coincidence $ \gamma $ spectra, gated by 766 keV to show the $ \gamma $ rays of present interest in \isotope[131]Xe. The $ \gamma $ rays marked in magenta colour are already placed in the positive parity band in this nucleus \cite{131Xe_PRC}.}
\label{spec-1}
\end{figure}

\section{Experimental details}

Excited states in \isotope[131]Xe were populated via \isotope[130]Te(\isotope[4]He, 3n$ \gamma $) fusion-evaporation reaction at $ E_{\alpha} = 38 $ MeV. Energetic beam of $ \alpha $ particles was delivered by the K-130 cyclotron of the Variable Energy Cyclotron Centre (VECC), Kolkata, India \cite{VECC_cyclotron}. Isotopically enriched target of \isotope[130]Te, evaporated on a 600 $ \mu $g/cm\textsuperscript{2} thick mylar backing, was 2 mg/cm\textsuperscript{2} thick. Seven Compton-suppressed clover HPGe detectors of the Indian National Gamma Array (INGA), arranged in three different angles $ \theta = 40^{\circ}, 90^{\circ}, 125^{\circ} $ with respect to the beam direction, were employed to detect the de-exciting $ \gamma $ rays \cite{VECC-INGA}. A Pixie-16 (250 MHz, 12 bit, 16 channel) based digital data acquisition system was utilised to record the time-stamped valid events in single and coincident modes \cite{DAQ_VECC_UGC-DAE-CSR_NIM-A}. The gain-matched raw data were sorted to construct the symmetric and asymmetric matrices using the {\scriptsize IUCPIX} \cite{DAQ_VECC_UGC-DAE-CSR_NIM-A} data sorting package. Offline data analysis was carried out using {\scriptsize INGASORT} \cite{INGAsort} and {\scriptsize RADWARE} \cite{RADWARE} analysis packages. Further details of this experiment and the data analysis procedures are available in Ref.~\cite{Ranabir_thesis,131Xe_PRC}.

\section{Results}

From the present data analysis, a cascade of three $ \gamma $-transitions with E$ _{\gamma} = $ 705, 816 and 930 keV has been established above the $ I^{\pi} = 9/2^{-} $ state (\figurename~\ref{lev-sch}). The levels of this sequence are also found to decay into the $ \alpha = -1/2 $ yrast negative parity levels through 177, 882, 1055 and 1176 keV $ \gamma $ rays. Pertinent energy gated $ \gamma $ spectra in favour of the new placements are shown in \figurename~\ref{spec}. In this work, the earlier reported 816 and 1055 keV $ \gamma $-transitions are found to be decaying from the same $ I^{\pi} = 17/2^{-} $ state at E\textsubscript{level} = 1862 keV \cite{131Xe_PRC,131Xe_PS}. Observation of both of these $ \gamma $-rays in the 930 keV energy gated spectrum [\figurename~\ref{spec}(b)] provide necessary support in favour of this assignment. The $ \gamma $ ray peak that is observed at 881 keV in the total projection of the E$ _{\gamma} $-E$ _{\gamma} $ matrix is found to be a doublet, composed of two transitions, of 880 keV and 882 keV. Thus, the placement of the 882 keV transition, in presence of the 880 keV doublet, was difficult due to the lack of sufficient gating transitions. It shows a clear centroid difference between the peaks present in the 663 and 816 keV energy-gated spectra [\textit{see}, \figurename~\ref{spec}(a) and \ref{spec}(c)]. Non-observation of the 882 keV transition in the coincidence spectrum of the 705 keV gate provides further support in favour of its placement (\figurename~\ref{spec}).
\begin{table}[!t]
\caption{Energies (E$ _{\gamma} $ and E\textsubscript{level}, in keV), relative intensities (I$ _{\gamma} $), DCO ratios (R\textsubscript{DCO}, extracted from stretched \textit{E}2 gates), linear polarization asymmetries ($ \Delta $\textsubscript{asym}) and spin/parity ($ I^{\pi} $) assignments for the $ \gamma $-rays/levels in \isotope[131]Xe.}
\label{results} 
\begin{ruledtabular}
\begin{tabular}{cccccc}
E$ _{\gamma} $&E\textsubscript{level}&I$ _{\gamma} $&R\textsubscript{DCO}&$ \Delta $\textsubscript{asym}&$ I^{\pi}_i \rightarrow I^{\pi}_f $	\\
\hline
177.3&341.3&3.4 (2)&0.57 (2)& &$ 9/2^{-} \rightarrow 11/2^{-} $	\\
188.8&1805.6&3.1 (2)&0.82 (2)&&$ 19/2^{+} \rightarrow 19/2^{-} $	\\
204.8&1805.6&$ < $ 1&0.63 (2)&&$ 19/2^{+} \rightarrow 17/2^{-} $	\\
389.0&2194.6&2.4 (1)&1.09 (2)&$ + $0.08 (2)&$ 23/2^{+} \rightarrow 19/2^{+} $	\\
444.0&2249.5&1.2 (1)&0.61 (1)&$ - $0.05 (3)&$ 21/2^{+} \rightarrow 19/2^{+} $	\\
642.3&806.0&100&1.01 (1)&$ + $0.08 (1)&$ 15/2^{-} \rightarrow 11/2^{-} $	\\
662.5&3180.9&11.7 (6)&0.97 (1)&$ + $0.12 (2)&$ 27/2^{-} \rightarrow 23/2^{-} $	\\
670.6&834.6&$ > 1.8 $&0.57 (4)&$ - $0.03 (5)&$ 13/2^{-} \rightarrow 11/2^{-} $	\\
704.6&1046.0	&2.0 (1)&1.28 (32)&$ + $0.02 (1)&$ 13/2^{-} \rightarrow 9/2^{-} $	\\
766.5&1600.9&$ \approx 1.8 $&1.15 (20)&&$ 17/2^{-} \rightarrow 13/2^{-} $	\\
794.8&1600.9&12.6 (2)&0.55 (1)&$ + $0.03 (1)&$ 17/2^{-} \rightarrow 15/2^{-} $	\\
810.5&1616.2&81 (4)&1.01 (1)&$ + $0.10 (2)&$ 19/2^{-} \rightarrow 15/2^{-} $	\\
815.6&1861.9&1.0 (1)&0.92 (7)&$ + $0.03 (6)&$ 17/2^{-} \rightarrow 13/2^{-} $	\\
880.1&4061.0&2.7 (1)&0.57 (3)&$ - $0.10 (3)&$ 29/2^{-} \rightarrow 27/2^{-} $	\\
881.8&1046.0&1.6 (1)&0.64 (11)&&$ 13/2^{-} \rightarrow 11/2^{-} $	\\
901.9&2517.8&21 (1)&1.01 (1)&$ + $0.10 (1)&$ 23/2^{-} \rightarrow 19/2^{-} $	\\
929.7&2792.1&$ < 1 $&0.91 (24)&&$ 21/2^{-} \rightarrow 17/2^{-} $	\\
990.0&3508.2&2.8 (2)&0.94 (3)&$ + $0.24 (3)&$ 27/2^{-} \rightarrow 23/2^{-} $	\\
1055.1&1861.9&6.0 (2)&0.54 (3)&$ + $0.02 (5)&$ 17/2^{-} \rightarrow 15/2^{-} $	\\
1175.9&2792.1&$ \approx 1 $&0.64 (7)&$ - $0.06 (9)&$ 21/2^{-} \rightarrow 19/2^{-} $	\\
\end{tabular}
\end{ruledtabular}
\textbf{Note.} 
Uncertainty in the E$ _{\gamma} $ is about 0.1 - 0.2 keV. 
The E\textsubscript{level} are determined by fitting the measured energies of the $ \gamma $-rays shown in the \figurename~\ref{lev-sch}.
The I$ _{\gamma} $, R\textsubscript{DCO} and $ \Delta $\textsubscript{asym} of the known $ \gamma $-rays are adopted from Ref.~\cite{131Xe_PRC}.
\end{table}

\begin{figure}[!b]
\centering
\includegraphics[width=0.495\columnwidth, valign=t]{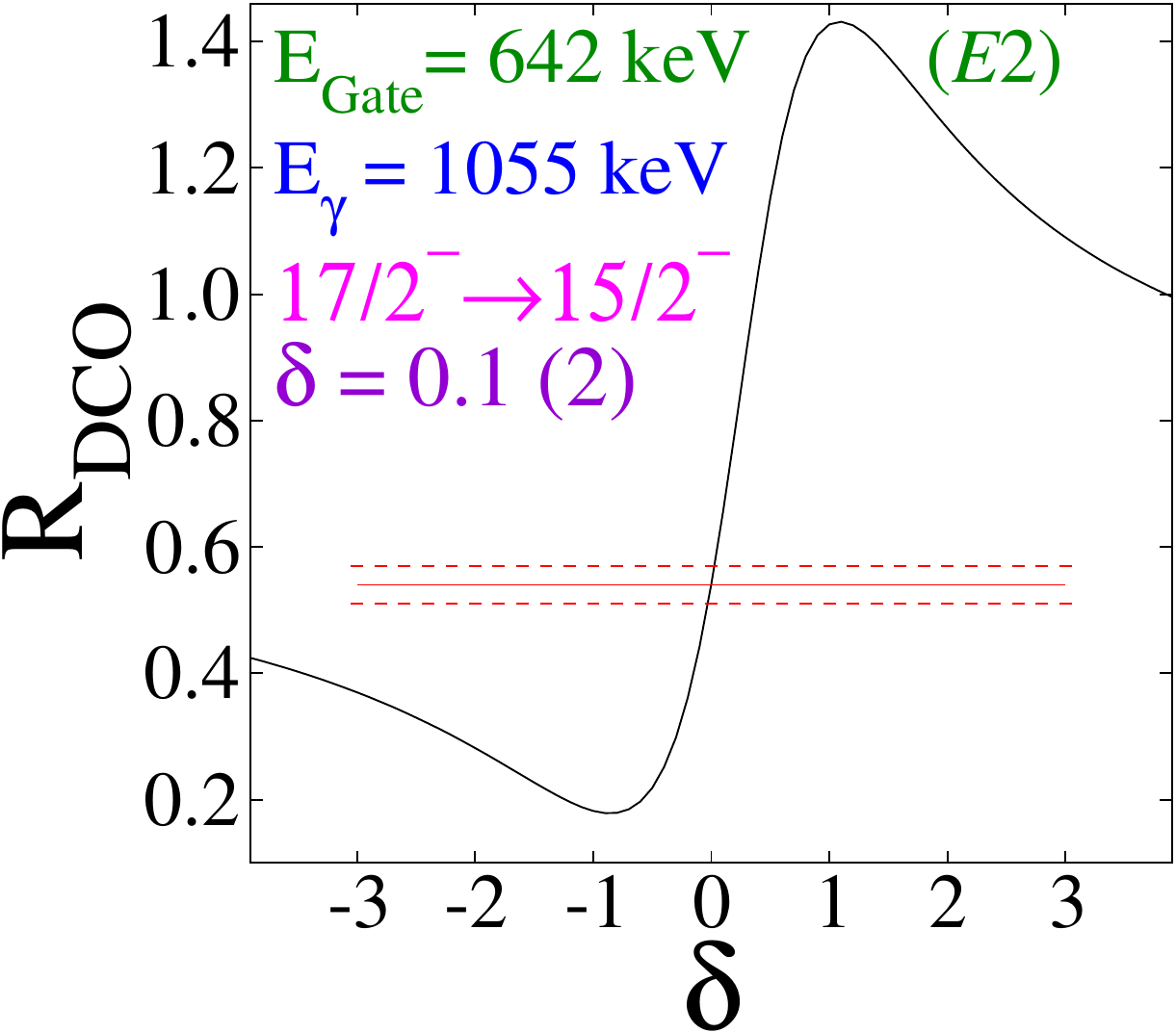}
\includegraphics[width=0.485\columnwidth, valign=t]{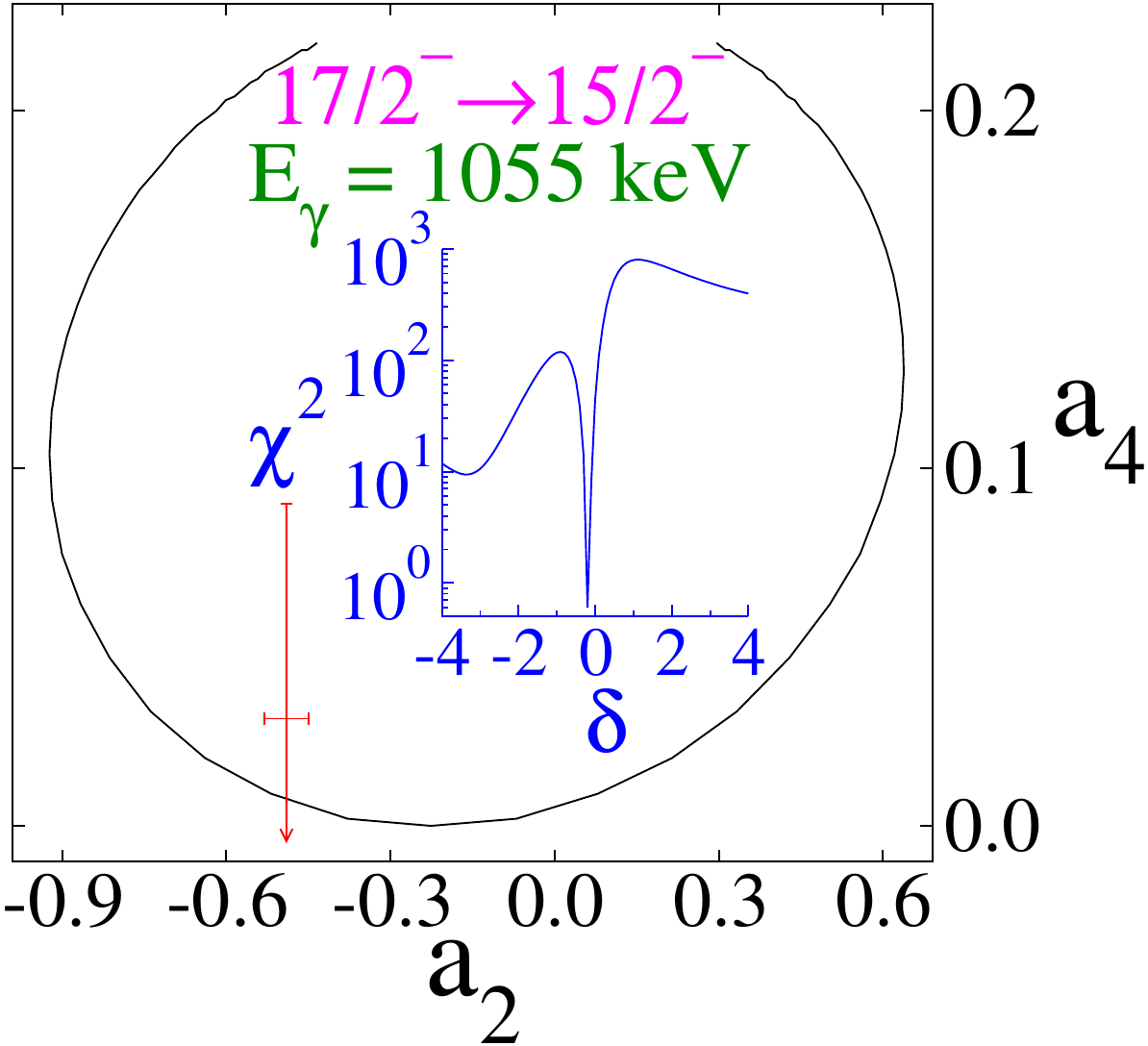}
\caption{Left: Variation of the theoretical DCO ratio as a function of multipole mixing ratio $ \delta $ (black line). The solid (dashed) red lines indicate the experimentally measured R\textsubscript{DCO} (error). Right: Contour plot of the angular distribution coefficients, a\textsubscript{2} versus a\textsubscript{4}, for different $ \delta $ (black line). Corresponding dispersion of experimental data, taken from Ref.~\cite{131Xe_PS}, is marked with red cross. The $ \chi^{2} $ analysis for experimental distribution is shown in the inset (blue).}
\label{ang_cor_dis}
\end{figure}

\begin{figure}[!t]
\centering
\includegraphics[width=\columnwidth]{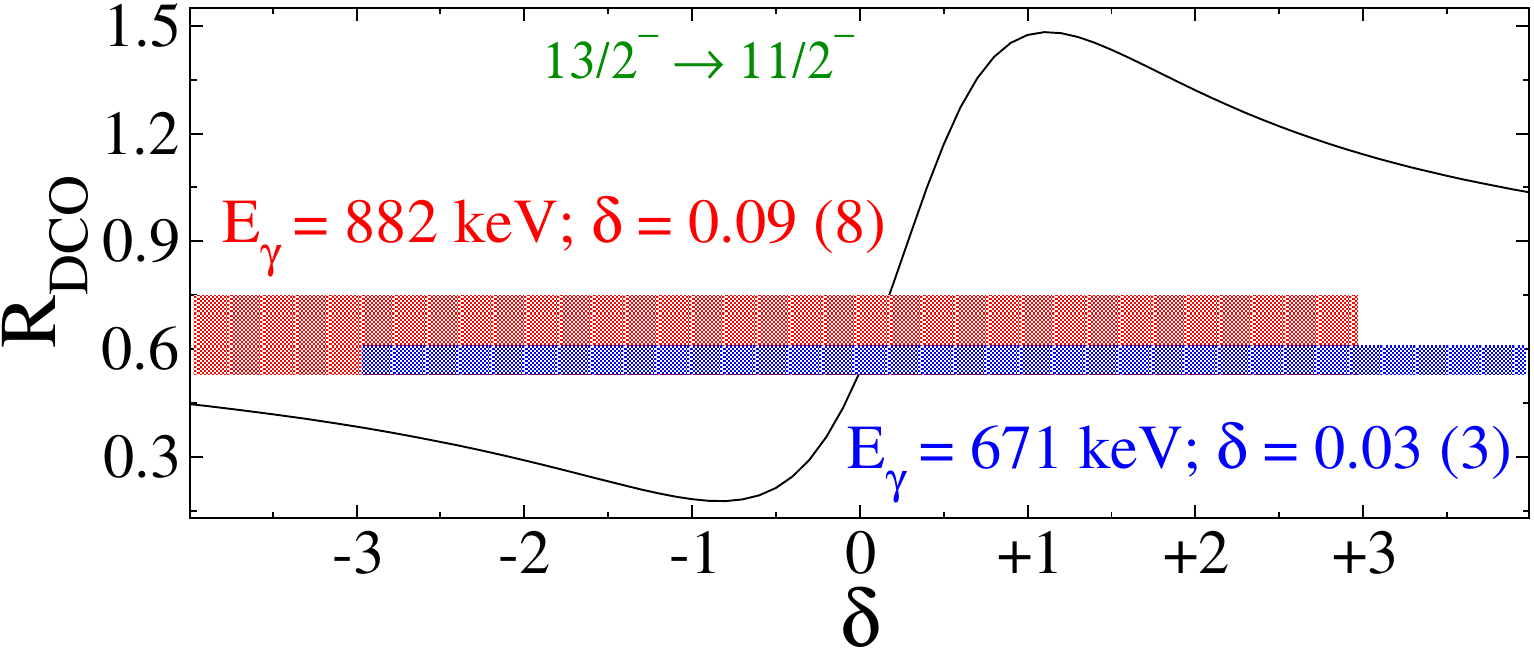}
\caption{Variation of the theoretical DCO ratio as a function of multipole mixing ratio $ \delta $ (black line) for the $ 13/2^{-} \rightarrow 11/2^{-} $ transitions. The red (blue) shaded region corresponds to the experimental DCO ratio of 882 (671) keV transition within the limits of uncertainty.}
\label{13/2to11/2}
\end{figure}
This sequence is extended further by placing an E$ _{\gamma} = $ 930 keV transition, based on its coincidence with the 705 and 816 keV $ \gamma $ rays (\figurename~\ref{spec}). However, the higher spin states of this band were not observed in the present work. Apart from these, two more $ \gamma $-rays, \textit{viz.}, E$ _{\gamma} = $ 671 and 766 keV, are also identified in this work. The E$ _{\gamma} = $ 766 keV transition is found to decay from the $ I^{\pi} = 17/2^{-} $ state at E\textsubscript{level} = 1601 keV to a newly established state at E\textsubscript{level} = 835 keV (\figurename~\ref{lev-sch}). This  835 keV state is observed to decay further to the isomeric $ I^{\pi} = 11/2^{-} $ state at E\textsubscript{level} = 164 keV through the newly identified transition of E$ _{\gamma} = $ 671 keV (\figurename~\ref{lev-sch}). The observation of the 671 keV transition and the absence of the 795 keV $ \gamma $-line in the 766 keV energy gated spectrum, as shown in \figurename~\ref{spec-1}, provide support for the placement of these new transitions.

To understand the origin of these newly observed sequences one needs to determine the multipole mixing ratio $ \delta $ of the connecting $ \gamma $-transitions. The $ \delta $ of a $ \gamma $-ray can be determined by comparing the experimentally measured angular distribution/correlation and/or linear polarization with their theoretical estimates. However, the linear polarization of weak transitions has large uncertainty and hence, the angular correlation/distribution is used mainly to determine the multipole mixing ratio. In this work, the multipolarities of the observed $ \gamma $ rays are determined from the angular correlation (R\textsubscript{DCO}) measurement. Experimentally deduced R\textsubscript{DCO} confirms the dipole character of the 671, 882, 1055 and 1176 keV $ \gamma $-rays and quadrupole character of the 705, 766, 816 and 930 keV $ \gamma $-rays, as listed in \tablename~\ref{results}. The $ E2/M1 $ multipole mixing ratios ($ \delta_{E2/M1} $) of the inter-band $ \Delta I = 1 $ $ \gamma $-rays have been estimated by comparing the experimental R\textsubscript{DCO} with its theoretical values. The theoretical values of R\textsubscript{DCO} for different $ \delta $ have been calculated using the computer code {\scriptsize ANGCOR} \cite{ANGCOR}. As a representative case, the variation of calculated DCO ratio, as a function of $ \delta $, has been shown in the left panel of \figurename~\ref{ang_cor_dis} for the 1055 keV ($ 17/2^{-} \rightarrow 15/2^{-} $) $ \gamma $-transition. The magnitude of $ \delta_{\text{1055 keV}} $ is found to be approximately zero (\figurename~\ref{ang_cor_dis}) from the present analysis which indicates almost pure $ M 1 $ character of the 1055 keV $ \gamma $ ray. In an earlier experiment \cite{131Xe_PS}, the angular distribution coefficients [a\textsubscript{2} = -0.49 (4) and a\textsubscript{4} = 0.03 (6)] of the 1055 keV $ \gamma $ ray were measured. The calculated a\textsubscript{2}-a\textsubscript{4} contour plot, along with these experimentally measured values, has been shown in the right panel of \figurename~\ref{ang_cor_dis}. The corresponding $ \chi^{2} $ analysis, as shown in the inset of the right panel of \figurename~\ref{ang_cor_dis}, also provides a low value of $ \delta $, which is in good agreement with the $ \delta $ value determined from the present angular correlation measurement. The similar magnitudes of R\textsubscript{DCO} = 0.55 (1), a\textsubscript{2} = -0.47 (3) and a\textsubscript{4} = -0.03 (8) are indicative of a similarly lower $ \delta_{E2/M1} $ for the E$ _{\gamma} = $ 795 keV $ \gamma $ transition from $ I^{\pi}_{i} = 17/2^{-} $ to $ I^{\pi}_{f} = 15/2^{-} $ \cite{131Xe_NPA}. From a similar comparison of experimental and calculated DCO ratio as shown in \figurename~\ref{13/2to11/2}, it is also evident that the magnitudes of $ \delta $ of the two newly identified $ \Delta I = 1 $ $ \gamma $ transitions between $ 13/2^{-} \rightarrow 11/2^{-} $ states (\textit{i.e.}, $ E_{\gamma} = 671,~882 $ keV) are also very low ($ \lessapprox 0.1 $) and thus have very little $ E2 $ ($ \approx 1\% $) admixture.
In addition, the electric or magnetic nature of the $ \gamma $ transitions have been determined from the linear polarization asymmetry ($ \Delta $\textsubscript{asym}) measurement. 

\section{Discussion}

\begin{figure}[!t]
\centering
\includegraphics[width=\columnwidth]{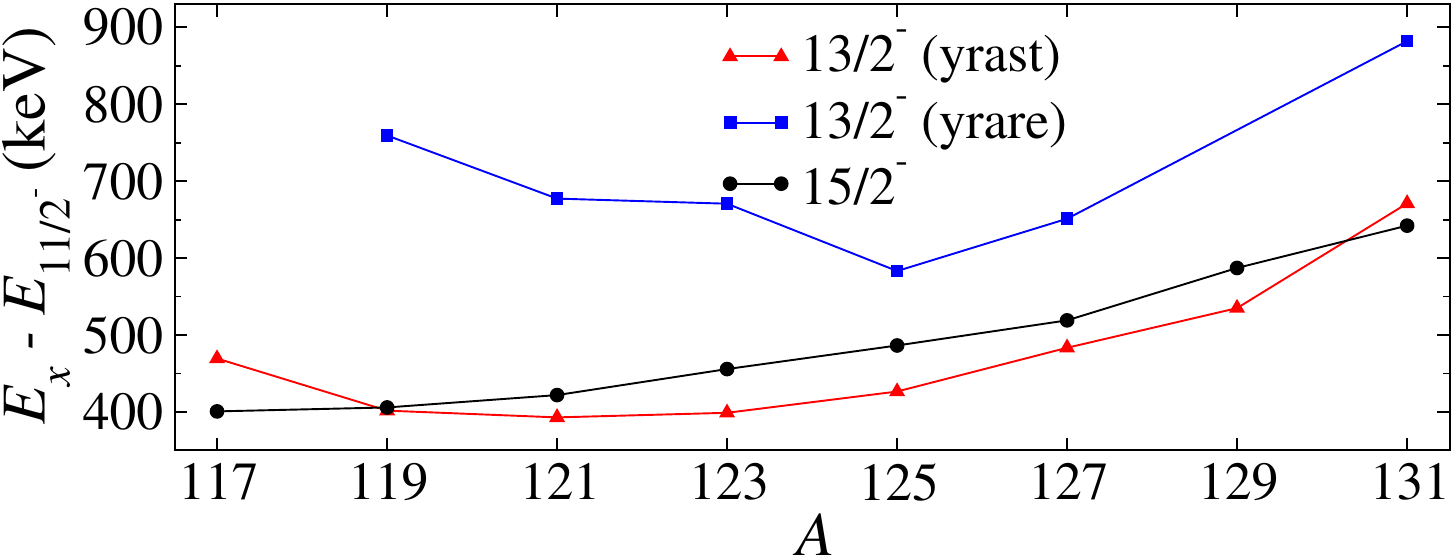}
\caption{Variation of the excitation energy of $ 15/2^{-} $ and $ 13/2^{-} $ (yrast and yrare) states, relative to the $ E_{11/2^{-}} $, in \textsuperscript{117--131}Xe, as a function of mass number (\textit{A}). Experimental data taken from Refs.~\cite{117Xe,119Xe,121Xe,123Xe,125Xe,127Xe_PLB,129Xe_PRC,131Xe_PRC}.}
\label{sys}
\end{figure}

\begin{figure*}[!t]
\centering
\includegraphics[width=\textwidth]{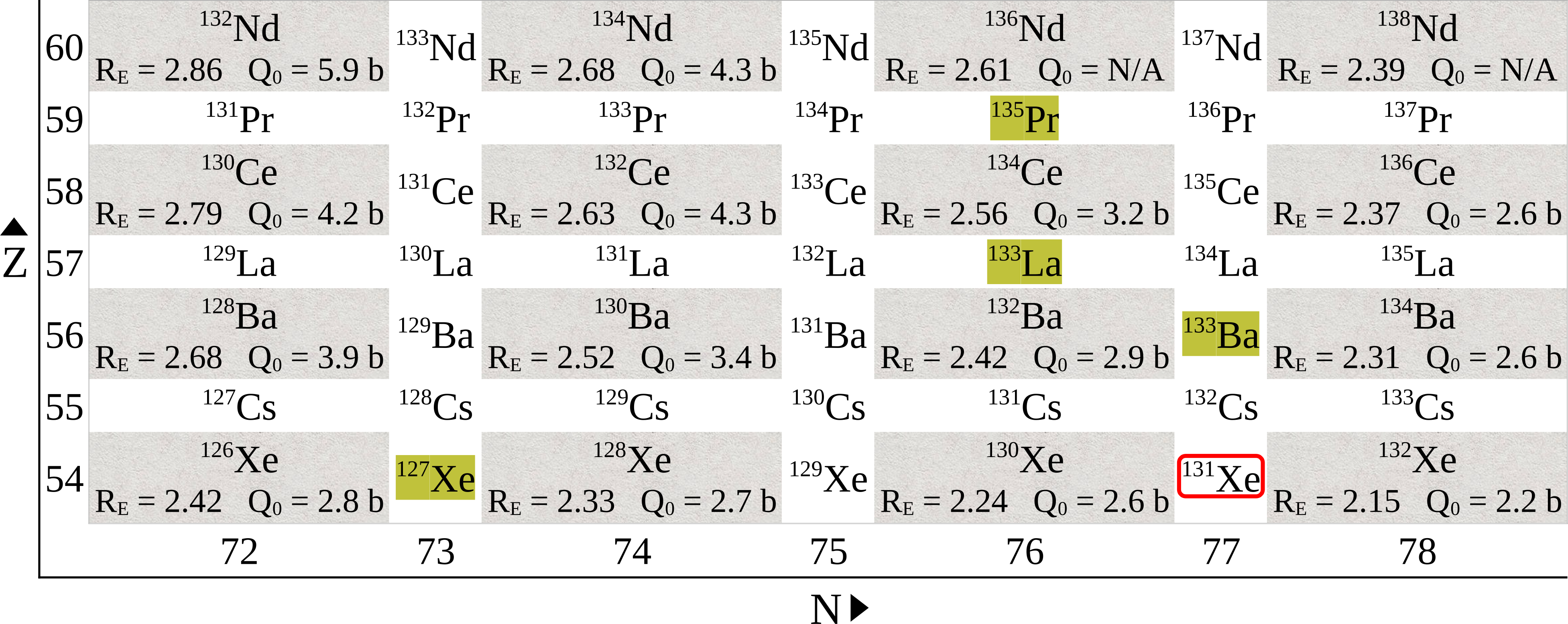}
\caption{A section of the Segr\`e chart showing the 4\textsuperscript{+} to 2\textsuperscript{+} energy ratio (R\textsubscript{E} = $ E_{4^{+}}/E_{2^{+}} $) and the intrinsic quadrupole moment (Q\textsubscript{0}) of the \textit{even-even} nuclei around $ A \approx 130 $ \cite{NNDC,Q0}. The highlighted odd-\textit{A} nuclei are known for their wobbly nature \cite{127Xe_PLB,133Ba_PLB,133La_EPJA,135Pr_PRL}. The nucleus of present interest, \isotope[131]Xe, is marked with a red border.}
\label{wobb_sys}
\end{figure*}

The effect of intruder $ h_{11/2} $ orbital, which gives an important piece of nuclear structural information in the proximity of $ A \approx 125 $ region, can be studied from the bands based on the $ \nu h_{11/2} $ orbitals in Xe-Ba isotopes. Being associated with mid/high-$ \Omega $ configurations, both the $ \alpha = \pm 1/2 $ signature partners of the $ \nu h_{11/2} $ band were expected to be observed in these nuclei. Systematically, two $ I^{\pi} = 13/2^{-} $ states were reported in \isotope[119-127]Xe \cite{119Xe,121Xe,123Xe,125Xe,127Xe_PLB}. The excitation energies of these $ I^{\pi} = 13/2^{-} $ and $ I^{\pi} = 15/2^{-} $ states relative to the energy of $ I^{\pi} = 11/2^{-} $ state (\textit{i.e.}, $ E_{x} $ after subtracting $ E_{11/2^{-}} $) are plotted as a function of mass number in \figurename~\ref{sys} for all the Xe isotopes with $ A = 117-131 $. It is evident from this figure that the yrast (yrare) $ 13/2^{-} $ state in \isotope[119-127]Xe is systematically found at a lower (higher) excitation energy than the $ 15/2^{-} $ state as shown in \figurename~\ref{sys}. In the case of \isotope[117,129]Xe, only the yrast $ I^{\pi} = 13/2^{-} $ state was reported \cite{117Xe,129Xe_PRC}. 
The rotational band built on the yrast $ 13/2^{-} $ states in \isotope[119-125]Xe and \isotope[129]Xe were designated as the unfavoured signature partner of the $ \nu h_{11/2} $ band. On the other hand, the band on yrare $ 13/2^{-} $ states in \isotope[119-125]Xe were interpreted as the quasi-$ \gamma $ band coupled to the $ h_{11/2} $ quasineutron \cite[and references therein]{125Xe_PRC}. However, these interpretations are not sufficient to justify the observed signature splitting in these bands, as discussed in Ref.~\cite{125Xe_PRC}. Recent works on \isotope[127]Xe ($ Z = 54 $, $ N = 73 $) \cite{127Xe_PLB} and \isotope[133]Ba ($ Z = 56 $, $ N = 77 $) \cite{133Ba_PLB} reveal that the $ 13/2^{-}_{\text{yrast}} \rightarrow 11/2^{-} $ $ \gamma $-transition is predominantly $ E2 $ ($ \Delta I = 1 $) in nature and based on this, a different interpretation in terms of \textit{wobbling motion} was proposed for the bands built on these states \cite{127Xe_PLB,133Ba_PLB}. In contrast to this, the $ 13/2^{-}_{\text{yrare}} \rightarrow 11/2^{-} $ $ \gamma $-transition is found to be primarily $ M1 $ in nature and hence, the corresponding band is considered as the unfavoured signature partner of the $ \nu h_{11/2} $ band. However, unlike lighter \isotope[117-129]Xe isotopes ($ Z = 54 $) and \isotope[133]Ba, \isotope[135]Ce isotones ($ N = 77 $), only the favoured signature partner ($ \alpha = +1/2 $) of the $ \nu h_{11/2} $ band was known from the previous studies on \isotope[131]Xe \cite{131Xe-PRC,131Xe_PRC}.

From the present study, both the yrast and yrare $ I^{\pi} = 13/2^{-} $ states in \isotope[131]Xe are identified. But, in contrast to \isotope[127]Xe and \isotope[133]Ba, the $ 13/2^{-}_{\text{yrast}} \rightarrow 11/2^{-} $ and $ 17/2^{-}_{\text{yrast}} \rightarrow 15/2^{-} $ $ \gamma $-transitions in \isotope[131]Xe are found to be predominantly $ M1 $ and thus excluding the possibility of wobbling nature of the sequence built on the yrast $ I^{\pi} = 13/2^{-} $ state. To infer the structural evolution behind such a change, the properties \textit{viz.}, 4\textsuperscript{+} to 2\textsuperscript{+} energy ratio (R\textsubscript{E} = $ E_{4^{+}}/E_{2^{+}} $) and the intrinsic quadrupole moment (Q\textsubscript{0}), of the even-even nuclei in this mass region have been revisited. From this study it is found that the odd-\textit{A} nuclei, where a wobbling band has been observed, are surrounded by two even-even nuclei with R\textsubscript{E} $ \geq $ 2.3 and Q\textsubscript{0} $ \geq $ 2.6 b as shown in \figurename~\ref{wobb_sys}. Also, two even-even nuclei, namely, \isotope[130]Ba \cite{130Ba_wobbling} and \isotope[136]Nd \cite{136Nd_wobbling}, having R\textsubscript{E} $ > $ 2.5 are reported to exhibit wobbling mode. From Casten's symmetry triangle it is known that the 4\textsuperscript{+} to 2\textsuperscript{+} energy ratio for a $ \gamma $-soft [O(6)] nuclei is found to be R\textsubscript{E} = 2.5 \cite{Casten_CPS}. However, the nucleus of current interest, \isotope[131]Xe, is surrounded by even-even \isotope[130,132]Xe nuclei with R\textsubscript{E} $ \approx $ 2.2 and Q\textsubscript{0} $ \leq $ 2.6 b. With R\textsubscript{E} $ \approx $ 2.2, the \isotope[130,132]Xe nuclei are found to be good candidates for Z(4) model and thus indicates a transition from vibrator [U(5)] to $ \gamma $-soft [O(6)] critical point symmetries \cite{Z_4-128-132Xe}. Therefore, it perhaps can provide an information about the effect of R\textsubscript{E} and/or Q\textsubscript{0} on the emergence of wobbling motion in atomic nuclei. However, the number of experimentally identified wobbling bands are quite limited and more experimental investigations on the odd-\textit{A} nuclei are required to understand the mechanism behind appearance or disappearance of wobbling mode in triaxial nuclei. 

\begin{figure}[!b]
\centering
\includegraphics[height=\columnwidth, angle=270]{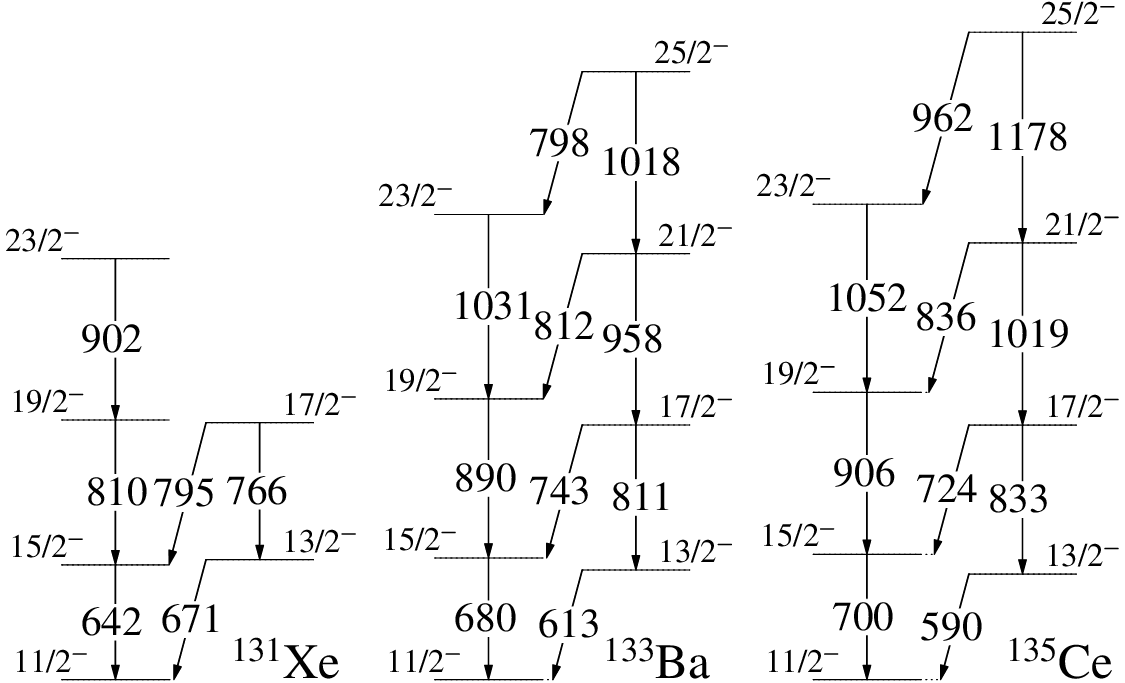}
\caption{Observed $ \Delta I = 2 $ bands on the yrast $ I^{\pi} = 13/2^{-} $ states in \isotope[131]Xe [this work], \isotope[133]Ba \cite{133Ba_PLB} and \isotope[135]Ce \cite{135Ce} $ N = 77 $ isotones.}
\label{wobb_sys_LS}
\end{figure}

\begin{figure}[!t]
\centering
\includegraphics[width=\columnwidth]{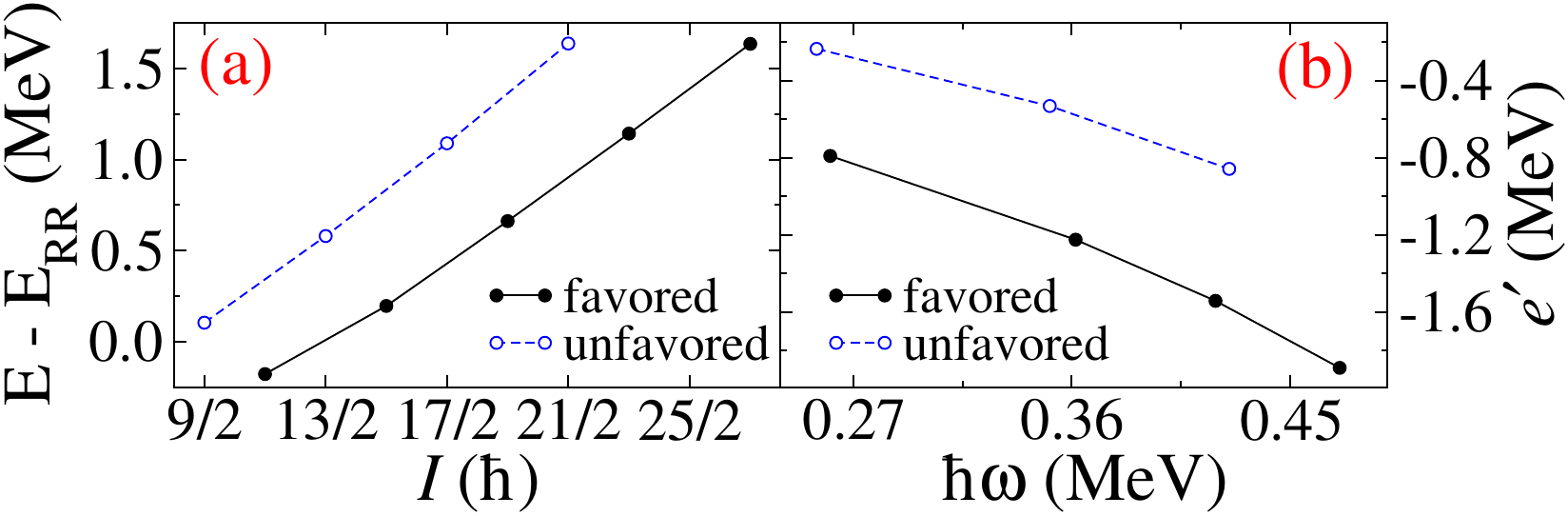}
\caption{Plots of (a) the observed level energies, after subtracting an average rigid-rotor energy, as a function of angular momentum and (b) the experimental Routhian as a function of rotational frequency for the $ \nu h_{11/2} $ band in \isotope[131]Xe.}
\label{E-e}
\end{figure}

\begin{figure}[!b]
\centering
\includegraphics[width=0.49\columnwidth]{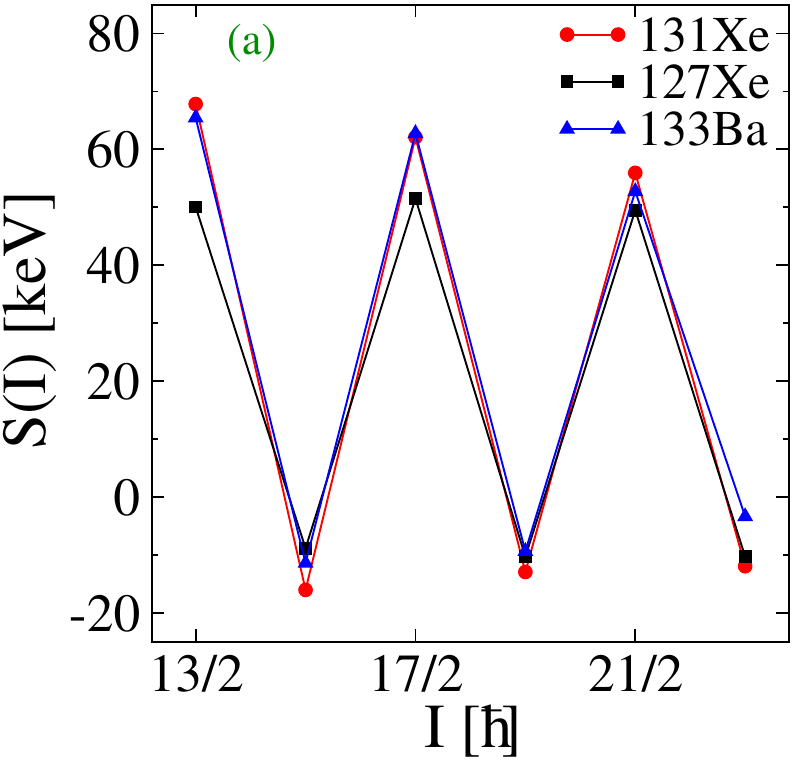}
\includegraphics[width=0.49\columnwidth]{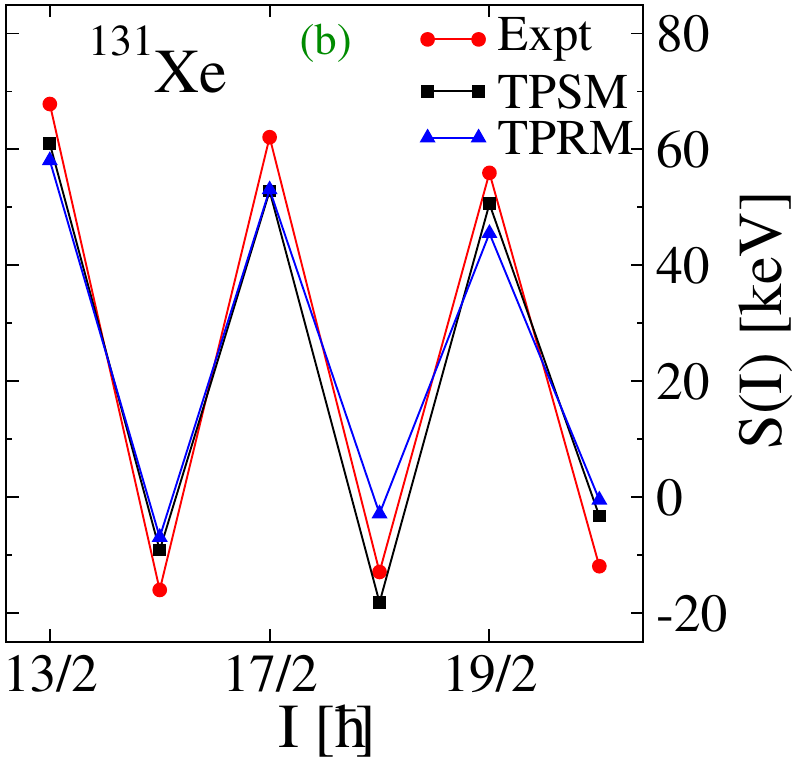}
\caption{(a) Left: Signature splitting of the $ \alpha = \pm 1/2 $ partners of $ \nu h_{11/2} $ band in \textsuperscript{127}Xe \cite{127Xe_PLB}, \textsuperscript{131}Xe [this work] and \isotope[133]Ba \cite{133Ba_PLB}. (b) Right: Comparison of the experimental energy staggering between favoured and unfavoured signature partners of $ \nu h_{11/2} $ band with its theoretical estimates for \textsuperscript{131}Xe.}
\label{sp}
\end{figure}

\begin{figure*}[!t]
\centering
\includegraphics[height=\textwidth, angle=270]{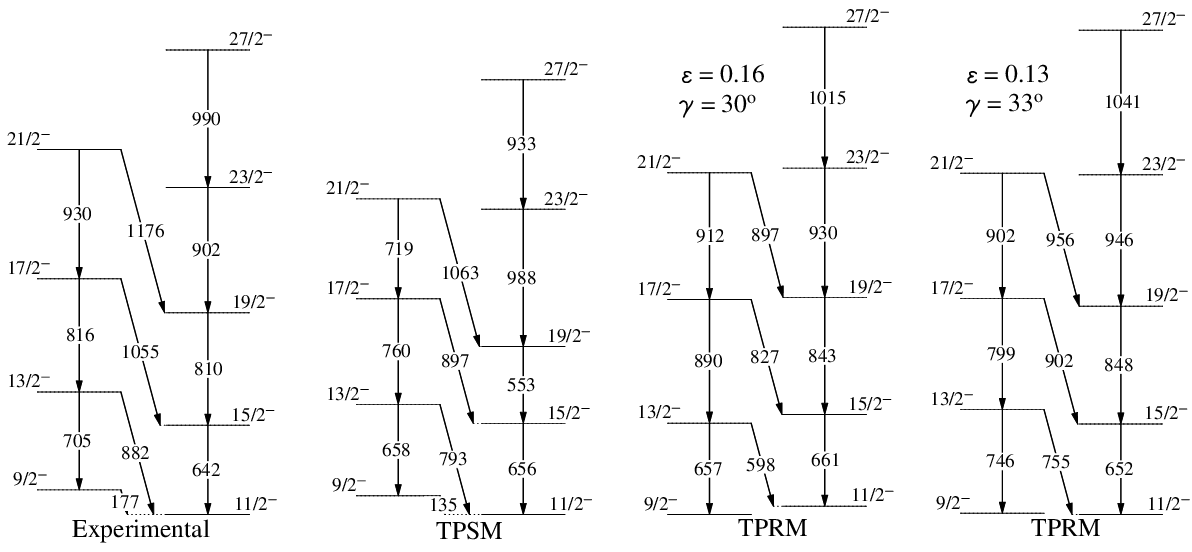}
\caption{A comparison of the experimental $ \nu h_{11/2} $ band structure in \isotope[131]Xe with the theoretical predictions of the triaxial projected shell model (TPSM~\cite{odd-Xe_TPSM}) and triaxial particle rotor model (TPRM~\cite[present work]{Xe_triaxiality_NPA}).}
\label{131Xe-LS}
\end{figure*}

To check the above correlation between wobbling motion and R\textsubscript{E} or Q\textsubscript{0}, the available spectroscopic information on the \isotope[135]Ce has been reinvestigated as a test case. The nucleus \isotope[135]Ce is of special interest as its neighbours show a variety of structural phenomena associated with the triaxial shapes of atomic nuclei. For instance, excitation of wobbling mode has been reported in \isotope[136,138]Nd \cite{136Nd_wobbling,138Nd_wobbling_chiral}, \isotope[135]Pr \cite{135Pr_PRL}, \isotope[134]Ce \cite{134Ce_wobbling}, \isotope[133]La \cite{133La_EPJA} and \isotope[133]Ba \cite{133Ba_PLB} nuclei. Another experimental fingerprint of non-axial nuclear shapes is the observation of a pair of chiral doublet bands. Evidence of chiral doublet bands have been found in \isotope[135]Ce \cite{135Ce_chiral}, as well as, in several other neighbouring nuclei, such as, \isotope[135-138]Nd \cite{135Nd_Chiral,135Nd-Chiral,136Nd_Chiral,137Nd_chiral,138Nd_wobbling_chiral}, \isotope[133]Ce \cite{133Ce_chiral}, \isotope[133]La \cite{133La_chiral} and \isotope[131]Ba \cite{131Ba_chiral}. Apart from these, a number of $ \gamma $-vibrational bands have been observed in this region. Recently, a new kind of excitation mode, namely, tilted precession (TiP), has also been reported in \isotope[135]Nd \cite{135Nd_TiP}. The magnitudes of R\textsubscript{E} and Q\textsubscript{0} of the neighbouring even-even \isotope[134,136]Ce nuclei are found suitable to observe the wobbling excitation in triaxially deformed \isotope[135]Ce (\figurename~\ref{wobb_sys}). \figurename~\ref{wobb_sys_LS} shows the negative parity bands in \isotope[131]Xe, \isotope[133]Ba and \isotope[135]Ce associated with $ \nu h_{11/2} $ configuration. The sequence above the $ I^{\pi} = 13/2^{-} $ state in \isotope[133]Ba is found to be due to the excitation of wobbling mode \cite{133Ba_PLB}. Similar to the case of \isotope[133]Ba, the magnitude of the DCO ratio of the connecting $ \Delta I = 1 $ transitions, \textit{e.g.}, $ E_{\gamma} = 590~\text{and}~724 $ keV, in \isotope[135]Ce are also found to be low compared to that expected for a pure dipole transition \cite{135Ce}. Also, the a\textsubscript{2} and/or a\textsubscript{4} angular distribution coefficients of these transitions are found higher than expected. These indicate a large $ E2 $ admixture in these $ \gamma $ rays and hence the band above the $ I^{\pi} = 13/2^{-} $ state in \isotope[135]Ce is also found to be a possible candidate for a wobbling band. Detailed experimental fingerprint for this expected wobbling band in \isotope[135]Ce can be explored in a future experiment.

The spin-parity assignment and the decay pattern of the newly established band above $ I^{\pi} = 9/2^{-} $ state in \isotope[131]Xe make it a suitable candidate for the unfavoured signature partner ($ \alpha = -1/2 $) of $ \nu h_{11/2} $ band. The multipole mixing ratio, estimated from the present spectroscopic results, indicates low \textit{E}2 admixture in the $ \Delta I = 1 $ inter-band transitions. Such a low magnitude of $ \delta $ is also found to be in agreement with this interpretation. The rotational properties like observed excitation energies relative to a rigid rotor (E-E\textsubscript{RR} \cite{E-ELD}) and experimental Routhian ($ e^{\prime} $ \cite{Routhian}) are also found to exhibit similar character (\figurename~\ref{E-e}) and hence providing further support in favour of the signature relationship of these bands. 

The right  panel of \figurename~\ref{sp} shows the signature splitting, as a function of spin, of the $ \alpha = \pm 1/2 $ partners of the $ \nu h_{11/2} $ band in \textsuperscript{127}Xe \cite{127Xe_PLB}, \textsuperscript{131}Xe [this work] and \isotope[133]Ba \cite{133Ba_PLB}. The signature splitting [$ S(I) $] of the $ \nu h_{11/2} $ band in \isotope[131]Xe is found to be as large as that reported in \isotope[127]Xe and \isotope[133]Ba. Such an unexpectedly large $ S(I) $ of the negative parity yrast levels was considered as a signature of $ \gamma $ deformation in \isotope[125]Xe \cite{125Xe_PRM}. Evidence of triaxial deformation in the form of wobbling motion was also reported in \isotope[127]Xe and \isotope[133]Ba \cite{133Ba_PLB,127Xe_PLB}. Thus, \isotope[131]Xe may also possess a triaxial nuclear shape. Theoretical calculations successfully reproduced the $ S(I) $ of the $ \nu h_{11/2} $ band in several odd-\textit{A} Xe-Ba isotopes \cite{Xe_triaxiality_NPA}.  Recently, a detailed theoretical investigation was carried out on \isotope[117-131]Xe under the framework of the extended triaxial projected shell model (TPSM) \cite{odd-Xe_TPSM}. This calculation is also able to reproduce the experimental $ S(I) $ of \isotope[131]Xe quite successfully [\figurename~\ref{sp}(b)], but, the calculated relative excitation energies of the states, \textit{i.e.}, $ \Delta $E\textsubscript{level} $ \equiv $ E$ _{\gamma} $, in some cases differ from the corresponding experimental values. For instance, the magnitude of the $ 19/2^{-} \rightarrow 15/2^{-} $ ($ 23/2^{-} \rightarrow 19/2^{-} $) transition was smaller (larger) than expected (\figurename~\ref{131Xe-LS}). In this work, a further theoretical investigation based on the triaxial particle rotor model (TPRM) is carried out to infer the microscopic structure of the negative parity band in \isotope[131]Xe.

\subsection*{Triaxial Particle Rotor Model calculations}

The triaxial particle rotor model is a widely used approach to study the spectroscopic features of low-lying states in the odd mass nuclei theoretically. Under this framework, an odd nucleon, located in a deformed single particle orbital, is supposed to rotate in the asymmetric potential of a triaxilly deformed even-even core, characterised by the quadrupole deformation parameters ($ \epsilon_{2}, \gamma $). The Hamiltonian ($ H $) of the coupled system is comprised of the collective rotation of the even-even core ($ H_{core} $), the single-particle Hamiltonian ($ H_{sp} $) and the pairing interaction ($ H_{pair} $). Thus, the total particle-rotor Hamiltonian can be expressed as:
$$ H = H_{core} + H_{sp} + H_{pair} $$
The Hamiltonian of the rigid rotor can be written as:
$$ H_{core} = \sum \limits_{k = 1}^{3} A_{k} (I_{k} - j_{k})^{2} $$
where, $ A_{k} $ are the hydrodynamical inertial parameters. The total angular momentum is expressed as $ I $ and the particle angular momentum as $ j $. The single-particle Hamiltonian comprises of a modified oscillator potential characterized by the deformation parameters: $ \epsilon_{2} $, $ \epsilon_{4} $ and $ \gamma $. The pairing term is introduced via a standard BCS procedure to replace the single-particle energies by the quasiparticle energies. A complete description of this model is available in Refs.~\cite{TPRM-1,TPRM-2}.
      
To start with, the input parameters reported in Ref.~\cite{Xe_triaxiality_NPA} are used to calculate the excitation energies of the negative parity states in \isotope[131]Xe. This nicely reproduces the favoured partner, but, underestimates the signature splitting (\figurename~\ref{131Xe-LS}). The signature splitting, however, is found very sensitive to the triaxial deformation parameter $ \gamma $, as was also demonstrated in the case of \isotope[125]Xe \cite{125Xe_PRM}. The relative excitation energy of the $ I^{\pi} = 9/2^{-} $ state is also found different from its experimental value. Therefore, to achieve a good description of the experimental data, $ \epsilon_{2} = 0.13 $ and $ \gamma = 33^{\circ} $ deformation parameters were adopted finally \cite{131Xe_TPRM}. The Coriolis attenuation factor $ \xi = 1 $ has been used in this calculation \cite{131Xe_TPRM}. These reproduce both favoured and unfavoured signature partners of the $ \nu h_{11/2} $ band nicely, as shown in \figurename~\ref{131Xe-LS}. The left panel of \figurename~\ref{sp} shows a comparison of the experimental energy staggering between favoured and unfavoured signature partners of the $ \nu h_{11/2} $ band with its theoretical estimates for \textsuperscript{131}Xe.  From the present investigation, it can be concluded that the \isotope[131]Xe nucleus has a triaxially deformed shape and the observed large signature splitting in negative parity band arises primarily due to this $ \gamma $ deformation. 

To shed further light on the nature of the bands above the yrast and yrare $ I^{\pi} = 13/2^{-} $ states, the ratio of the reduced magnetic dipole and electric quadrupole transition probabilities [$ B(M1)/B(E2) $] has also been estimated. The calculated $ B(M1)/B(E2) = 1.27 $ is found in good agreement with the experimentally determined $ B(M1)/B(E2) = 1.27 (14) $ for the yrare $ I^{\pi} = 17/2^{-} $ state at E\textsubscript{level} = 1862 keV. The multipole mixing ratio, $ \delta_{\text{TPRM}} = -0.69 $, of the corresponding $ \Delta I = 1 $ transition ($ 17/2^{-} \rightarrow 15/2^{-} $) is also found to be low. On the other hand, the experimental $ B(M1)/B(E2) = 2.54 (4) $ of the yrast $ I^{\pi} = 17/2^{-} $ state is found to be two times higher in magnitude than that estimated from the TPRM calculation. Therefore, the band on the yrare $ 13/2^{-} $ state is considered as a more suitable candidate for the unfavoured signature partner of the $ \nu h_{11/2} $ band in \isotope[131]Xe. However, the origin of the yrast $ 13/2^{-} $ state still remains an open question and demands further investigations.

\section{Summary}

In-beam $ \gamma $ ray spectroscopy of \isotope[131]Xe has been carried out to study the structure of the intruder $ \nu h_{11/2} $ band utilizing the Indian National Gamma Array. Excited states were populated via $ \alpha $ induced fusion-evaporation reaction at E$ _{\alpha} = 38 $ MeV using the K-130 cyclotron of Variable Energy Cyclotron Centre, Kolkata, India. Inspection of $ \gamma \gamma $-coincidence data resulted in an identification of a rotational sequence which is found to decay into the yrast negative parity band. Based on the present spectroscopic results and the systematics of excitation energy, this sequence is proposed as the unfavoured signature partner of the $ \nu h_{11/2} $ band. The structure of this band is further investigated under the framework of the triaxial particle rotor model and the experimental outcomes are found in good agreement with theoretical estimates. Although the identification of the unfavoured signature partner of the $ \nu h_{11/2} $ band with a large signature splitting is a step forward towards the existence of the triaxial shape in this nucleus, no experimental signatures of the wobbling excitation have been found in the present study. Thus, it can be concluded that the presence of triaxial deformation alone perhaps does not warrant an observation of wobbling motion in atomic nuclei.

\section*{Acknowledgement}

The authors thankful to the INGA collaboration, partially funded by Department of Science and Technology, Government of India \textit{vide} Project No. IR/S2/PF-03/2003-I, for making the detectors available and setting up them at VECC, Kolkata. Effort from the Accelerator group (VECC) personnels is thankfully acknowledged for providing a quality $ \alpha $ beam during the experiment. Financial support of CEFIPRA/IFCPAR under Project No. 5604-4 is duly acknowledged. The first author (SC) sincerely acknowledges the financial support received from the Science and Engineering Research Board, Govt. of India (SERB, India) under National Post-Doctoral Fellowship (N-PDF) scheme \textit{vide} reference no. PDF/2022/001829. 

\bibliographystyle{apsrev4-2}
\bibliography{131Xe}

\end{document}